*Research Article*

# IoT-Based Water Quality Assessment System for Industrial Waste WaterHealthcare Perspective

**Abdur Rab Dhruba ,**[1] **Kazi Nabiul Alam ,**[1] **Md. Shakib Khan ,**[1] **Sananda Saha,**[1] **Mohammad Monirujjaman Khan ,**[1] **Mohammed Baz ,**[2] **Mehedi Masud ,**[3] **and Mohammed A. AlZain**[4]

[1]*Department of Electrical and Computer Engineering, North South University, Bashundhara, Dhaka 1229, Bangladesh*
[2]*Department of Computer Engineering, College of Computers and Information Technology, Taif University, P. O. Box 11099, Taif 21944, Saudi Arabia*
[3]*Department of Computer Science, College of Computers and Information Technology, Taif University, P. O. Box 11099, Taif 21944, Saudi Arabia*
[4]*Department of Information Technology, College of Computers and Information Technology, Taif University, P. O. Box 11099, Taif 21944, Saudi Arabia*

Correspondence should be addressed to Mohammad Monirujjaman Khan; monirujjaman.khan@northsouth.edu

The environment, especially water, gets polluted due to industrialization and urbanization. Pollution due to industrialization and urbanization has harmful effects on both the environment and the lives on Earth. This polluted water can cause food poisoning, diarrhea, short-term gastrointestinal problems, respiratory diseases, skin problems, and other serious health complications. In a developing country like Bangladesh, where ready-made garments sector is one of the major sources of the total Gross Domestic Product (GDP), most of the wastes released from the garment factories are dumped into the nearest rivers or canals. Hence, the quality of the water of these bodies become very incompatible for the living beings, and so, it has become one of the major threats to the environment and human health. In addition, the amount of fish in the rivers and canals in Bangladesh is decreasing day by day as a result of water pollution. Therefore, to save fish and other water animals and the environment, we need to monitor the quality of the water and find out the reasons for the pollution. Real-time monitoring of the quality of water is vital for controlling water pollution. Most of the approaches for controlling water pollution are mainly biological and lab-based, which takes a lot of time and resources. To address this issue, we developed an Internet of Things (IoT)-based real-time water quality monitoring system, integrated with a mobile application. The proposed system in this research measures some of the most important indexes of water, including the potential of hydrogen (pH), total dissolved solids (TDS), and turbidity, and temperature of water. The proposed system results will be very helpful in saving the environment, and thus, improving the health of living creatures on Earth.

## 1. Introduction

Polluted water causes different kinds of problems for human beings, animals, plants, and all other living beings. Urbanization and industrialization are increasing enormously with the proportionate rate of population growth [1]. Environmental pollution is a major concern in the era of industrialization. Water pollution is one component of this pollution. Pollution of water can cause severe damage to our health, impair the fertility of the land, and harm the aquatic life [2]. Industrial waste materials are one of the prime causes of water pollution. Textile dyeing is the second severest cause for water pollution globally, and the fashion industry produces 20% of the world's wastewater. In the manufacturing of clothing and household appliances, manufacturers' extensive water consumption and sanitation activities pollute soil and groundwater which negatively impact the environment, affecting ecosystems, and animal and human food chains. About 72 toxic chemicals can get mixed in the waterway while dyeing the fabric. These chemicals harm water and the entire ecosystem, exposing us to harmful chemicals and making us vulnerable to the resultant health hazards [3].

Recently, Bangladesh has achieved impetuous economic growth, and its target is to become a middle-income country by 2021. This sector comprises about 45% of our workforce [2]. The textile industry is an excellent locomotive for promoting the economic development of Bangladesh. The garments sector contributes to the 82% of the country's total export revenues, that is, about $28 billion a year. Currently, more than 3000 garment factories are located in Dhaka, according to the government database. The estimated wastewater emitted from these garment industries was about 217 million liters, and by the end of 2021, it will reach about 349 million liters. Industrial wastes and wastewater contain heavy metals such as vanadium, molybdenum, zinc, nickel, mercury, lead, copper, chromium, cadmium, and arsenic. This polluted water is used for irrigation in cultivated lands, paddy fields, vegetable growing fields, and other farming sources near industrial areas. Food poisoning, diarrhea, short-term gastrointestinal problems, respiratory diseases, skin problems, and some other severe health complications can be caused by this polluted water. It may have long-term consequences for our bodies [4].

So we can clearly understand that the textile industry is such a sector that has a long-term impact on our economy. Day by day, industrialization is taking place in different cities and areas to meet the needs of our society and our economy. But the expansion of these industries is harmful nowadays to our environment and, more specifically, our water. For these reasons, we can neither ignore the contribution of these industries nor stop their production. Hence, we need to find a solution to mitigate the risk factors by characterizing the water quality parameters. For this, we chose to measure the pH, temperature, turbidity, and TDS of water.

In search of some techniques for finding out the values of water quality parameters in the literature, we found in [5] that the authors used a Raspberry Pi and a YSI 600 module to measure these using a web scraping method. We found the use of IoT devices that controlled the basic Raspberry Pi and some sensors only in [6, 7]. In [8], we found an advanced approach for water quality measurement [8]. Here, the researchers designed a package system built with a controller and other sensors, using some different and alternative sensing modules that can monitor and show real-time data. They designed alternative sorts of flow sensors, conductivity sensors, temperature sensors, and pH sensors. Collectively, the sensors form a system that can gather data from water sources via a Zigbee receiver module to show an audio and visual representation of values. Computer Vision and Neural Networking approaches are also integrated to monitor water quality. As water quality classification models, the Recurrent Neural Network (RNN) and Long Short-Term Memory (LSTM) neural network techniques are used [9].

In [10], machine learning was aligned with IoT devices to sense and analyze water quality factors. We found the use of Arduino mkr100 module in [11], which is an old approach; the instrument is costly and currently less available in the market. We also found the use of STM32 module in measuring water quality parameters [12]. Some basic approaches related to our work have been found in [13, 14], as they added extra value to the monitoring system with their floating feature with a device that can be sent to rivers or water sources to get real-time data. Measuring and providing safe drinking water was the goal of articles [15–18]. They used a basic IoT approach with either a Wi-Fi or Bluetooth module, and an Arduino, Pi, or a combination of both. The primary goal of [19] was to measure and analyze water parameters in rural areas. A pulse monitoring system based on Arduino was proposed in [20]. In this study, to measure the heart rate of humans, the authors proposed only one sensor. The authors in [17] took measurements of body temperatures and heart rate, and displayed the results on a website platform. Numerous studies for IoT-based solutions in various domains have been conducted such as in [21, 22], and [23].

Most of the articles mentioned above used traditional ways to monitor water quality, and some of them showed different approaches in the monitoring. Some of the studies reported in these articles used advanced techniques like Neural Networking and Machine Learning approaches to analyze the parameters, and these studies mostly did so for monitoring drinking water or water sources that we use generally. But we found that they paid less importance or ignored assessing industrial water or, precisely, water released from textile industries. So, based on the discussion, we are motivated to propose an IoT device that will measure the water quality factors for industrial wastewater.

In this research, our main goal is to identify the harmful particles released in the wastewater from the textile industry. The proposed device/module detects solids, particles, gases, and other molecules dissolved into or separated from the wastewater emitted from the textile industry. Using our system, industry authorities can easily reduce the use of these elements in their chemicals or other materials that are mandatory in producing garment items. In this case, we fixed fundamental parameters for measuring water quality: pH, TDS, turbidity, and temperature. We chose those parameters to detect the waste water quality, as these parameters can show the toxic level of any kind of water. These four parameters almost cover the major factors in determining the quality of water. We also developed an IoT-based system which shows the real-time output data to recognize these parameters.

The following is a breakdown of the study's structure: The methods and methodologies for the study including outline of the system and details of the hardware materials, sensors, and working flowchart are discussed in Section 2. Section 3 focuses on the results of the study based on the analyses of the real-time data from the sensors. Finally, the research is summarized and the importance of safe water for living beings is discussed in Section 4, which finally concludes with future direction for this work.

## 2. Methods and Methodologies

*2.1. Methodology.* The methods, components, and procedures employed to attain the aim are discussed in this section. The system aims to measure the water quality of the industrial wastewater to control the pollution occurring due to the wastewater. The proposed microcontroller-based water quality monitoring system for industrial water comprises three different layers. The input and output layers are connected by the main layer, which is a microcontroller unit. The input layer is made up of four separate sensors that give an analog signal to the Arduino in order to measure various water quality indicators. The output layer consists of two parts: the microcontroller's monitor and a mobile application that displays the microcontroller's digital data conversion.

*2.2. Outline of System.* Figure 1 shows the block diagram of the system. Input, output, and an Arduino UNO microcontroller board make up the system. The Arduino board, which is also attached to the output units, is used in conjunction with the Arduino board's serial monitor and an MIT App inventor-based mobile application attached to a Bluetooth module to display the viewer digitally converted data.

Sensors pass analog data to the microcontroller Arduino UNO. The Arduino converts that data to a readable digital format and passes it to the mobile application through the Bluetooth module. It also displays it on the serial monitor of the Arduino.

*2.3. Materials.* The system consists of different kinds of components that perform different functions. Some are for the input and output, and some are used to bridge the input and output.

*2.3.1. Arduino UNO.* The main component of the system is the Arduino UNO REV3 based on the VR microcontroller Atmega328 with 2 kB SRAM and 32 kB of flash memory, of which 13 kB is used to store the instructions set in the form of code, and it has 1 kB of Electronically Erasable Programmable Read-only memory (EEPROM). Figure 2 shows the Arduino UNO pin out.

This Arduino board has 30 pins, with 14 digital pins and 6 analog pins for connecting to the outside environment. The A0 to A5 analog pins are used to receive analog data from external devices such as analog sensors. Several I/O (input/output) digital and analog pins are placed on the board which operates at 5 V. These pins come with standard operating ratings ranging from 20 mA to 40 mA. The 5 V voltage can be supplied by the Universal Serial Bus (USB) connection to an external device or the Direct Current (DC) power Jack, giving a voltage ranging from 7 V to 20 V. Serial communication is carried out through Pin 0 Receive (Rx) and Pin 1Transmit (Tx). The Rx pin is used to receive data, while the Tx pin is used to transmit data. Other I/O can be handy for serial communication. The Arduino UNO is a programmable microcontroller with the ability to interface with other sensors or computers. The Arduino Integrated Development Environment (IDE) software's serial monitor is used to send and receive text data from the device. Its purpose is to show the Arduino board's output data.

*2.3.2. Sensors.* There are several indexes to measure the quality of industrial water waste, including Total Dissolved Solids (TDS), Dissolved Oxygen, Photosynthetic Active Radiation (PAR), Life Science Parameters, Chlorophyll, Blue-Green Algae, ammonia, Ammonium, Biochemical Oxygen Demand (BOD), Turbidity, pH, and temperature. [22]. Among all these systems, the indexes in this study measure four of the most important parameters.

*(1) pH Sensor.* One index is the pH of the water. pH is the quantitative measure of the acidity or basicity of liquid solutions. Figure 3 shows the pH value of different kinds of solutions. The measurement of pH is conducted in a range from 0 to 14. The lower the pH of a solution is, the more acidic the solution is. High values of pH represent the basicity or alkalinity of a solution. The base solution usually has a very high pH of 10 to 14. Acidic solutions usually have a very low pH of 1–5. The ideal range of pH for water usually ranges from 6 to 8. Drinking water has a pH of 7.4. If the pH value of the water becomes very high or very low, it becomes toxic for the water animals and plants. As a result of this, the photosynthesis rate of water plants decreased. This hampered the health of water animals.

The pH of any solution is calculated by keeping the negative logarithm of hydrogen ion activity in the solution. Equation (1) shows the way to calculate the pH of any solution:

$$pH = -\log[H^+]. \quad (1)$$

The value of pH varies from one solution to another because of the hydrogen ion activity of the solution.

To measure the pH value of wastewater, the system contains a gravity pH sensor, Logo–Rnaenoar V2.0, as shown in Figure 4. This sensor is connected to the Arduino Uno board with an analog pin of the Arduino. This sensor passes analog data to the Arduino sensor by measuring the pH value of the wastewater. The Arduino converts the analog data into digital data to show it to the user. Figure 5 shows the pH sensor pinout.

This sensor returns a voltage proportional to the tendency of the solution to gain or lose electrons from other substances, which is linked directly to the pH of a substance. The sensor is powered up at a voltage of 5 V, and its working current is 5 mA–10 mA. One of the Arduino's analog pins is connected to the probe's P0 pin.

*(2) TDS Sensor.* Another index that the system include is the total dissolved solids (TDS). It represents the total concentration of dissolved substances in water. TDS can be measured using gravimetric analysis and conductivity. The gravimetric analysis approach is more appropriate for a solution where the salt level is relatively high. Conductivity is also useful for measuring the TDS of any solution. Equation

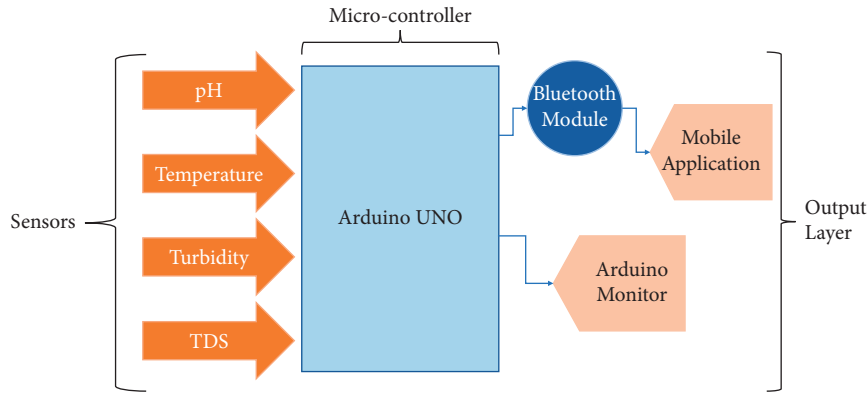

Figure 1: Block diagram of the entire system.

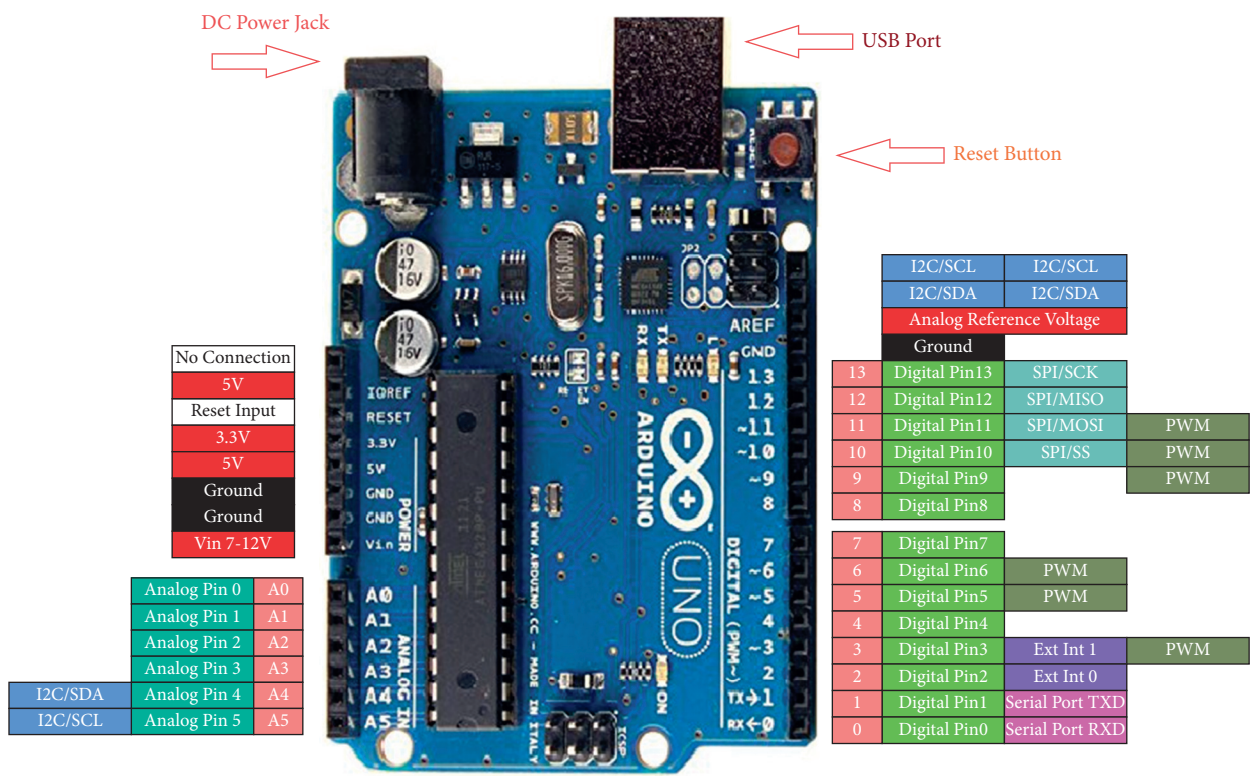

Figure 2: Arduino UNO pin-out [24].

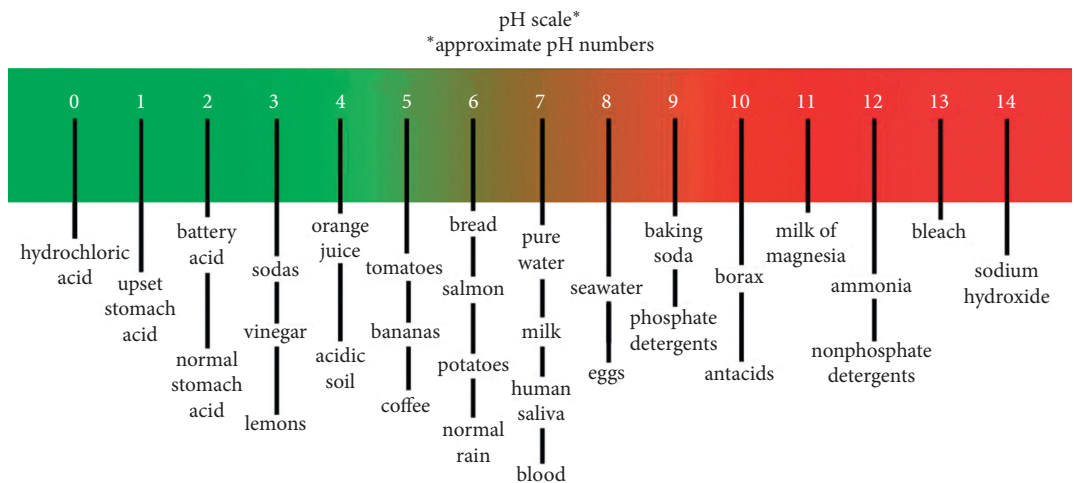

Figure 3: The pH value of different kinds of solution [25].

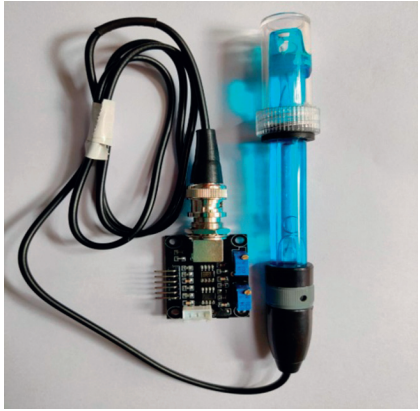

Figure 4: Gravity pH sensor.

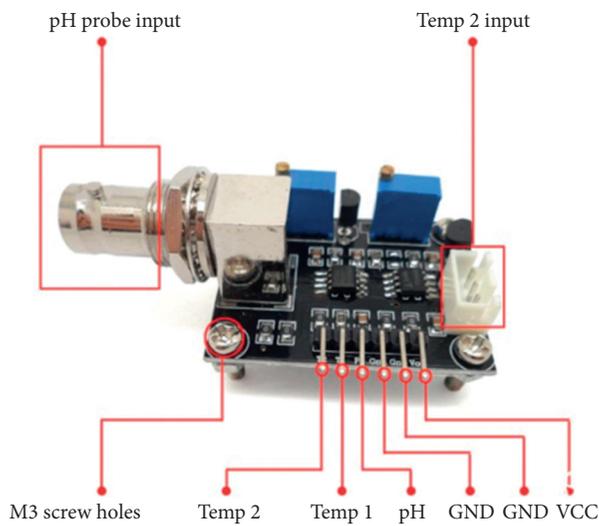

Figure 5: pH sensor pin out [26].

(2) can display the relationship between TDS and the specific conductance of water:

$$\text{TDS} = k_e \text{EC}. \tag{2}$$

Here, TDS is expressed in mg/L, and EC is the electrical conductivity in microsiemens per centimeter at 25°C. The correlation factor "ke" varies from 0.55 to 0.8. Figure 6 [27] shows the TDS range for different kinds of water.

This is an Arduino-supported e TDS meter kit to measure the TDS value of water which will determine the cleanliness of water. To measure the value of TDS, the system contains a sensor named gravity: Analog TDS Sensor V1.0. Figure 7 shows the TDS sensor that will pass an analog signal to the Arduino to convert it into digital data so that it can be displayed to the user.

The sensor's input voltage is between 3.3 V and 5.5 V, and its working current is between 3 mA and 6 mA. This board is easy to use with the Arduino boards. It passes the analog data to the Arduino through its output pin.

*2.3.3. Turbidity Sensor.* The system also measures the turbidity of the wastewater. Turbidity is the cloudiness or haziness of any fluid. It determines how much the water has lost its transparency as a result of the presence of suspended particles. It tells you how clean any fluid is. It is an optical characteristic of any fluid. The more suspended solids in the fluid and the cloudier the solution, the higher the turbidity is. The suspended particles in the water absorb heat from the sunlight, making the water reduce the oxygen level of the water. As a result of this, it harms the water creatures. Turbidity is measured in units named Nephelometric Turbidity Units (NTU). The turbidity values for different ranges are shown in Table 1.

Clean drinking water has a lower turbidity level. To measure the turbidity level of the wastewater, the system contains a turbidity sensor module kit to detect the cloudiness of the wastewater. The turbidity sensor, shown in Figure 8, sends an analog signal to the Arduino, which converts it to digital data that is displayed to the user.

The analog input is passed to the Arduino via this sensor, which is connected to the Arduino. The sensor's input voltage is 5 V, and its working current is a maximum of 30 mA. The probe of this sensor contains eight pins. The Arduino is attached to the data pin (*D*) of one of them. The Arduino gets the analog data through the data pin to convert it into digital output for the user.

*2.3.4. Water Temperature Sensor.* The system also measures the wastewater temperature to monitor the perfect temperature for the water animals to survive. The water temperature tells you how hot or cold the water is. It is an essential quality index for water as it also controls or affects some other indexes. The high temperature of the water is very harmful for fish. It reduces the rate of photosynthesis and prevents the growth of zooplankton, which is very important for fish. High temperatures reduce the oxygen level of the water and make the water saltier and more toxic.

To measure the waste water temperature, the system contains the LM-35 waterproof temperature sensor, as seen in Figure 9.

It is an Arduino-supported analog sensor used to measure temperatures in wet environments precisely. This sensor's data pin is connected to one of the Arduino UNO's analog pins. The LM35 is useful because the output voltage is linearly proportional to the centigrade temperature. This sensor can measure temperatures in a range of −55°C to 125°C. This sensor will work at a voltage level of 3 V to 5 V.

*2.3.5. MIT App Inventor 2.* A mobile application is included in the system to display real-time digital data from the Arduino board. MIT App Inventor 2 [29] was used to construct the mobile app. It is initially a web application integrated development environment provided by Google. The mobile application created is connected to the Arduino with a Bluetooth module. The layout of the MIT App inventor is shown in Figure 10.

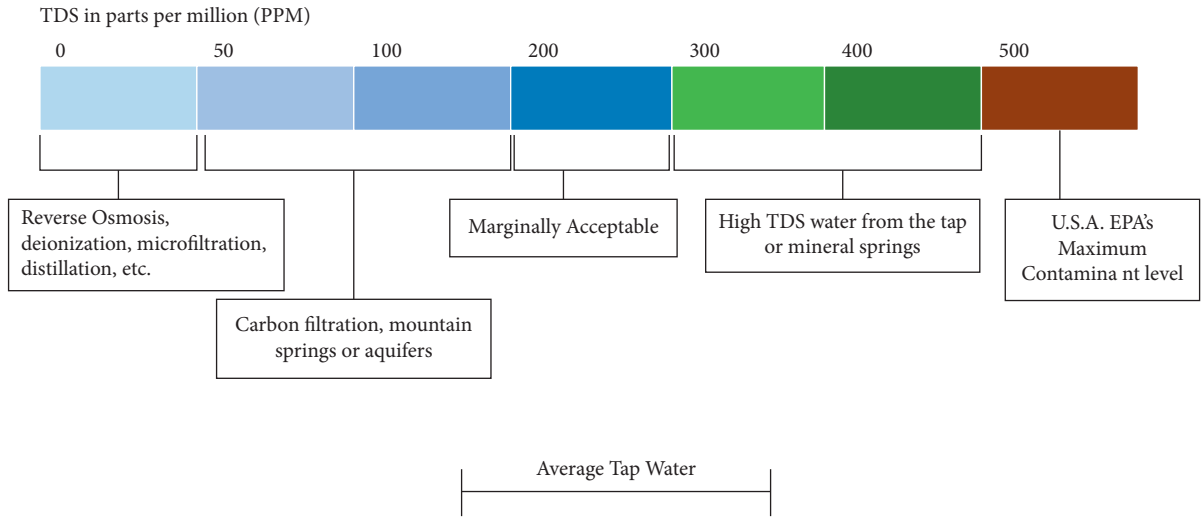

Figure 6: TDS value for different kinds of water [28].

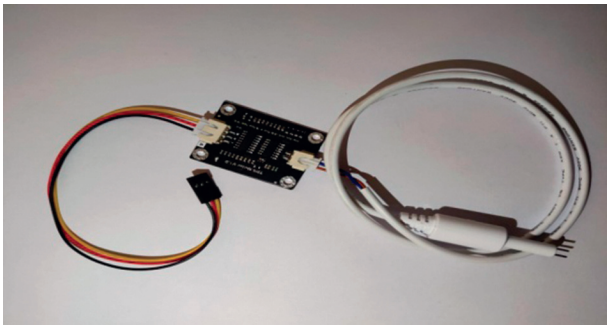

Figure 7: TDS sensor.

Table 1: Turbidity level with different turbidity values.

| No. | Turbidity level | TSM (NTU) |
| --- | --- | --- |
| 1 | Medium turbid | 0–25 |
| 2 | Rather turbid | 25–35 |
| 3 | Moderate turbid | 35–50 |
| 4 | Highly turbid | >50 |

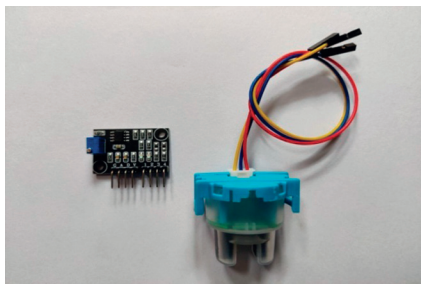

Figure 8: Water turbidity sensor module.

This web application was used to generate a mobile application for the system. As a bridge between the Arduino and the mobile application, a Bluetooth HC-05 module is linked to the Arduino. It has the functionality to add a two-way wireless connection.

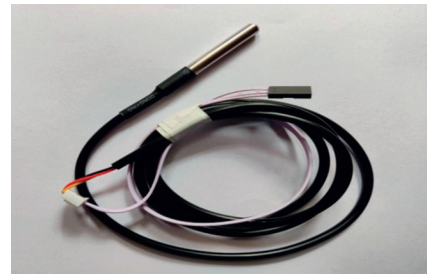

Figure 9: LM-35 Waterproof temperature sensor.

*2.3.6. Bluetooth Module.* This Bluetooth module has two operating modes. One is to send to or receive data from another device, and the other is to change the default device settings in AT Command mode. Users can see the digital data in the mobile application by pairing with this Bluetooth module easily. Figure 11 shows the pinout of the HC-05 Bluetooth module.

This module works at an operating voltage of 4 V–6 V and an operating current of 30 mA. The TX pin of this module works to transmit serial data. Data sent using this module passes through this pin. This pin is connected to the RX (Receive) pin of the Arduino, and the RX pin of the HC-05 works to receive serial data from the Arduino. The module's RX pin is connected to the Arduino's TX pin. After establishing a Bluetooth connection with the module, this module sends digital data from the Arduino to the mobile application.

*2.3.7. Working Process Flow-Chart.* The system comprises different steps. Every step has a separate task to do, and after all of the stages are completed, the user will find it useful. These several processes are combined in the programmable Arduino UNO software. The Arduino program's flowchart is displayed in Figure 12. The Arduino software brings together the various functionality for the various sensors and Bluetooth modules that connect the Arduino UNO to the mobile application.

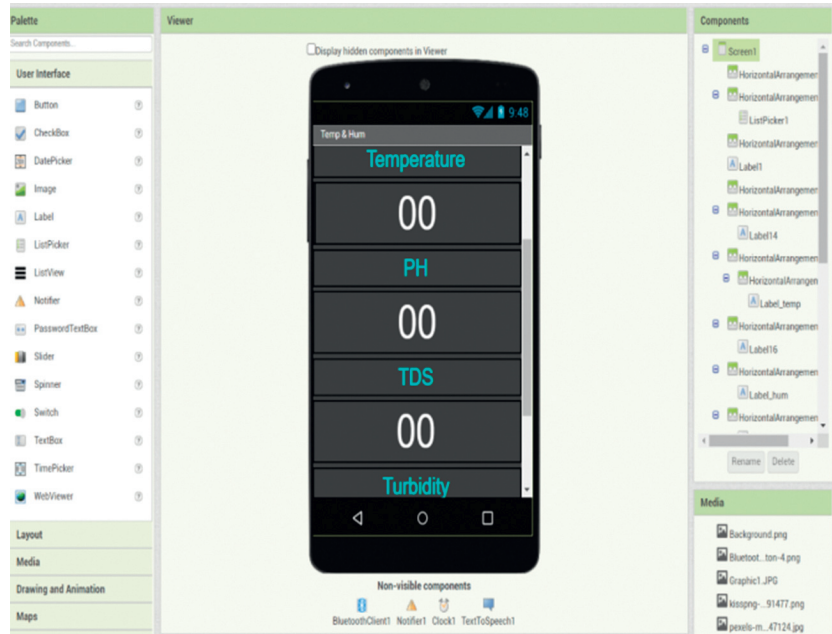

Figure 10: Layout of our App using MIT app inventor.

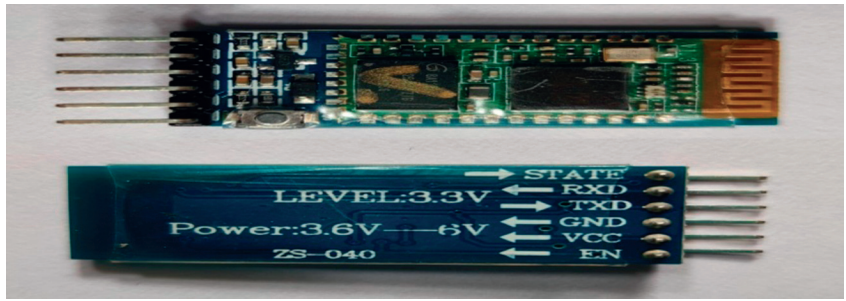

Figure 11: Bluetooth module.

Here, temperature, pH, TDS, and turbidity sensors sense the analog output from the water source and send their output analog signals to the microcontroller. The Arduino UNO receives those analog signals as input and converts them to digital signals which show in real-time values in the Arduino IDE monitor. A Bluetooth device is connected in between the two devices which pass this data to the mobile application showing the real-time fluctuations of water quality parameters.

## 3. Result and Analysis

*3.1. List of Sites Visited for Sample Collection.* We visited different sites to collect wastewater from different textile industries in Dhaka City in Bangladesh. We chose Savar and Gazipur where most of the garment industries are located. Table 2 shows the lists of sites from where we collected the water samples for this study.

*3.2. Design Prototype and View of Real-Time Data.* Our IoT device measured data for different water parameters, particularly wastewater, and the results are given here. After adding all the components, we uploaded the code to the Arduino board and measured data for all parameters, and it showed perfectly in our mobile application. Figure 13 shows the prototype of the developed system, including wastewater from different garment industrial sites. Figure 14 shows a mobile app showing different temperatures, pH, TDS, and turbidity parameters. This data are measured by the system we developed and shown on the mobile app in real-time.

### 3.3. Data Sheet and Analytical Charts for Different Measurements

*3.3.1. Temperature Analysis.* Table 3 shows the measured temperature values for different wastewater samples taken from different textile industries. Wastewater samples were collected from four different sites. For each site, five samples were collected and tested. Finally, the temperature values measured at each site were averaged. From the table, it is noted that different sites show different water temperature values. The value of the temperature of wastewater differs from factory to factory. Although in most cases, temperature values were in the range of 25–29°C, which is quite close to normal water temperature, there was one sample in the experiment where the temperature of wastewater was quite high reaching around 40°C. From the temperature data analysis, it appears that the water from the fourth site has high-temperature values in comparison with the water sample collected from other three sites. This may indicate that the wastewater from this site is more contaminated which needs to be considered as a crucial finding from this study.

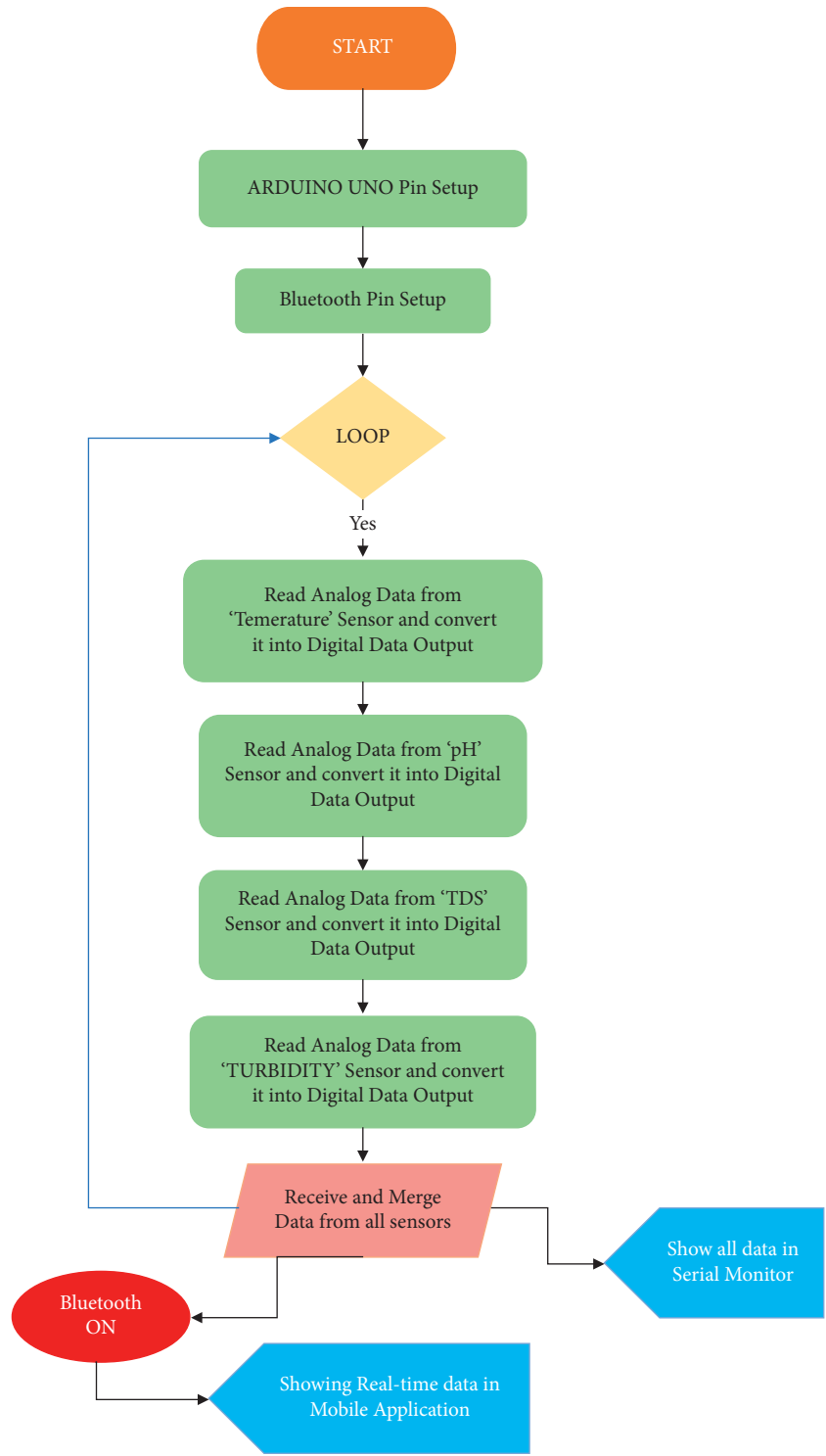

Figure 12: Flow-chart of the Arduino program.

Table 2: List of sites for sample collection.

| List of sites | Visited sites |
| --- | --- |
| Site-1 | GazipurSadar |
| Site-2 | SavarHemayetpur |
| Site-3 | GazipurKonabari |
| Site-4 | SavarExport processing zone (EPZ) |

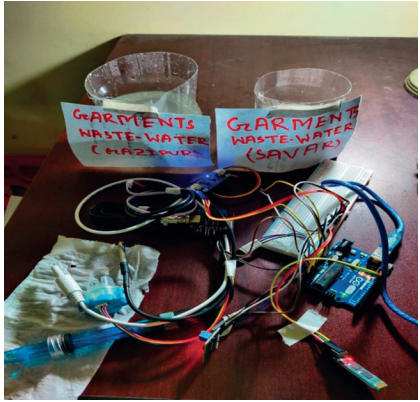

Figure 13: Prototype of the system.

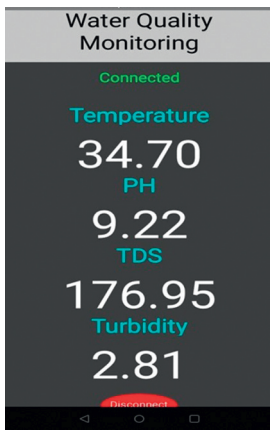

Figure 14: Mobile App showing real-time data.

Table 3: Temperature values for different samples.

| Temperature | Site-1 | Site-2 | Site-3 | Site-4 |
| --- | --- | --- | --- | --- |
| Take-1 | 28.84 | 26.88 | 23.46 | 36.17 |
| Take-2 | 29.81 | 32.26 | 24.93 | 39.10 |
| Take-3 | 32.26 | 28.35 | 22.97 | 40.57 |
| Take-4 | 25.90 | 27.73 | 28.35 | 33.34 |
| Take-5 | 27.86 | 24.21 | 25.90 | 38.12 |
| Average | 28.93 | 27.89 | 25.12 | 37.46 |

The water temperatures for the different samples collected from the various sites have been represented in a line chart in Figure 15. Site four is the industrial area located in Savar EPZ which is very near to Dhaka. The temperature of wastewater for site four is higher than the wastewater sample collected from the other three sites. To save the environment and the aquatic lives, this has to be taken into serious consideration by the proper authorities; otherwise, this wastewater will be a threat to not only the environment and the aquatic lives including fish and other beings living in the water but also human health.

*3.3.2. pH Analysis.* Table 4 lists the pH values or acidity and alkalinity values for different wastewater samples collected from various textile industrial sites. In this case, the sample was collected from the same locations as the ones from where the samples for measuring water temperatures were collected. From Table 4, it is noted that, in all cases, most samples have high pH values. All the samples have a pH value of more than 8.0 which tells us that all the samples contain molecules high in base solution. The standard pH level for healthy water is between 6 and 8, but in all cases, the pH values of the waste water released from the garments factories are higher, which is alarming. Compared with the four sites, the highest pH value is noticed in the water from site 2, while the lowest is noticed at site 4. Site 2 is located at Hemayetput in Savar-in Bangladesh. More details about the sites and water pH levels are depicted in Figure 16. Figure 16 represents pH values for all sites using a line graph in different colors. Site 4 in the graph shows the Savar EPZ area. Although site four has lower pH values than other sites, the values are also higher than the standard value which is very harmful for humans and other species living in the water.

The pH value in our experiment varied from industry to industry. Site 1 is the place where most of the garments industries use light-dyeing. The pH value of the wastewater released from this kind of industry is in the range of 9.00–9.90. Site 2 is the location where the garments factories use heavy dyeing, and where the pH values of the wastewater are in the range of around 10.50. At Site-3, most industries produce sportswear where the pH values of the wastewater are in the range of around 9.50, and lastly, at Site-4, most industries produce knit denim where the pH values of the wastewater change from 8.25 to 9.00.

*3.3.3. Turbidity Analysis.* Turbidity tells us how cloudy or hazy any solution is. Table 5 shows the turbidity values of different samples collected from the same four locations. The values of turbidity varied in a range of 1.30–1.60 for the three samples. This range is quite similar to that of normal tap water. But for one sample which is from Site 4, the turbidity value is higher than that in the other samples, but it is still at the fairly turbid level as shown in the turbidity table. From Table 5, it is observed that the highest turbidity value is noticed in the water collected from site 4 which is Savar EPZ. Figure 17 presents the turbidity values for all sites using a line graph. Here, turbidity values for samples collected from different locations have been presented in various colors. The lowest turbidity which was 1.30 was noticed in the waste water sample collected from Site 3.

The turbidity value of the waste water samples collected from Sites 1 and 4 is higher than those collected from Sites 2 and 3. At Site 1, most of the industries are light-dyeing industries but use heavily colored chemicals that cause the high turbidity in the wastewater. At Site 4, the color used in

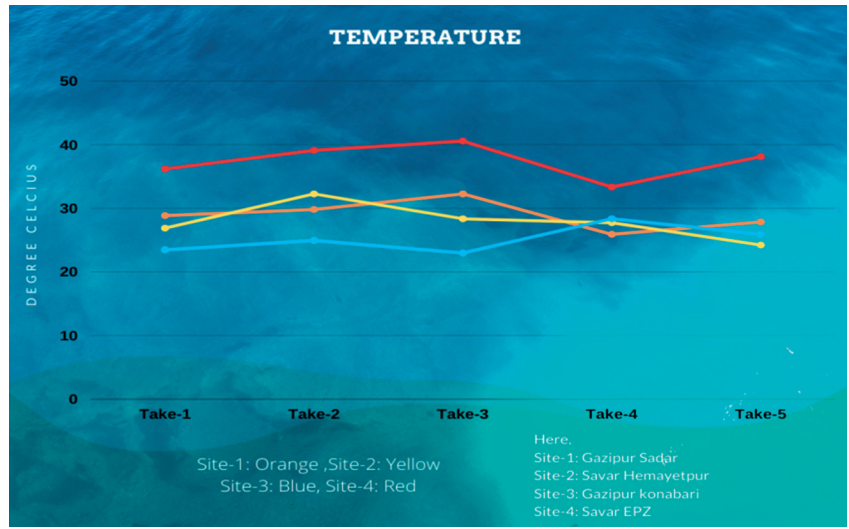

Figure 15: Line chart of temperature values for different wastewater samples from different sites.

Table 4: pH values for different waste water collected from different textile industries site.

| pH | Site-1 | Site-2 | Site-3 | Site-4 |
| --- | --- | --- | --- | --- |
| Take-1 | 9.08 | 10.57 | 9.06 | 8.27 |
| Take-2 | 9.48 | 10.65 | 9.27 | 8.41 |
| Take-3 | 9.66 | 10.61 | 9.39 | 8.62 |
| Take-4 | 9.76 | 10.42 | 9.34 | 8.98 |
| Take-5 | 9.86 | 10.56 | 9.42 | 8.73 |
| Average | 9.57 | 10.56 | 9.30 | 8.40 |

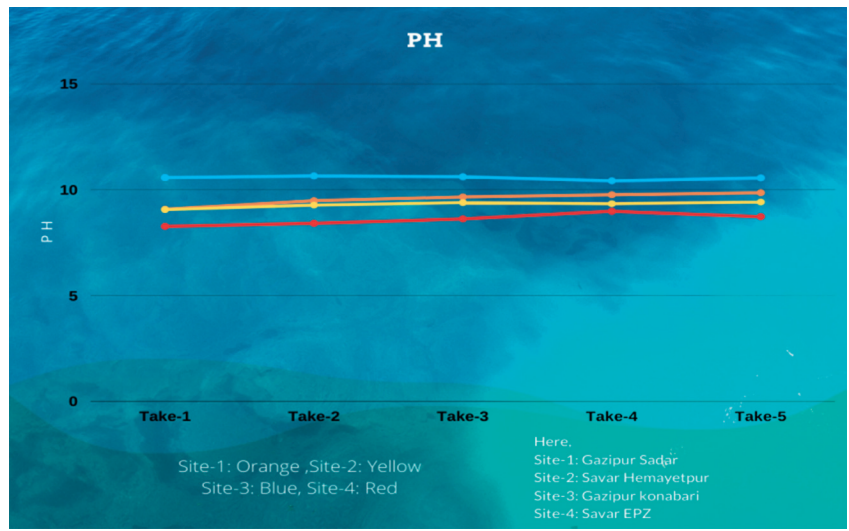

Figure 16: Line chart of pH values for different waste water samples.

Table 5: Turbidity values for different samples.

| Turbidity | Site-1 | Site-2 | Site-3 | Site-4 |
| --- | --- | --- | --- | --- |
| Take-1 | 2.00 | 1.33 | 1.29 | 2.94 |
| Take-2 | 1.95 | 1.36 | 1.30 | 2.95 |
| Take-3 | 1.97 | 1.31 | 1.32 | 2.98 |
| Take-4 | 1.94 | 1.42 | 1.31 | 2.88 |
| Take-5 | 1.87 | 1.38 | 1.30 | 2.89 |
| Average | 1.54 | 1.36 | 1.30 | 2.93 |

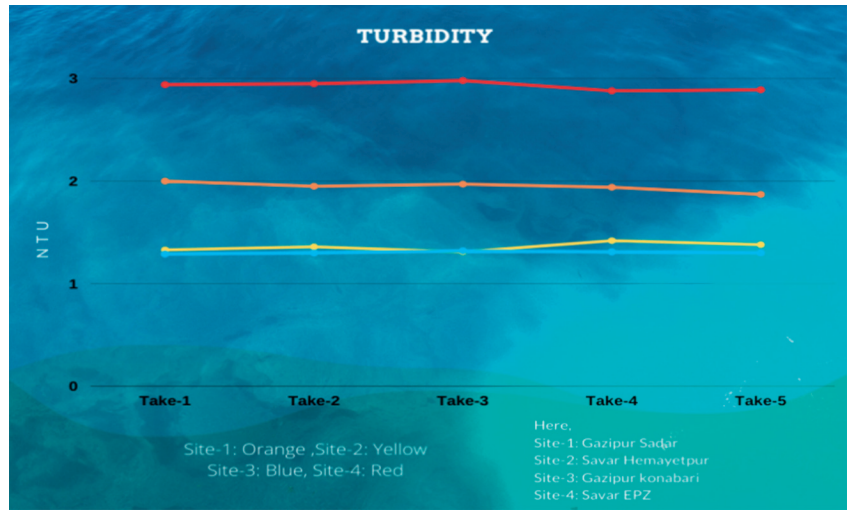

Figure 17: Line chart of turbidity values for wastewater samples collected from different sites.

Table 6: TDS values for different samples collected from different sites.

| TDS | Site-1 | Site-2 | Site-3 | Site-4 |
| --- | --- | --- | --- | --- |
| Take-1 | 349.50 | 187.36 | 352 | 170.12 |
| Take-2 | 348.23 | 185.47 | 356.95 | 172.06 |
| Take-3 | 349.50 | 183.58 | 353.25 | 185.47 |
| Take-4 | 350.75 | 189.25 | 353.25 | 175.92 |
| Take-5 | 350.75 | 185.47 | 358.17 | 177.85 |
| Average | 349.75 | 186.23 | 354.72 | 176.28 |

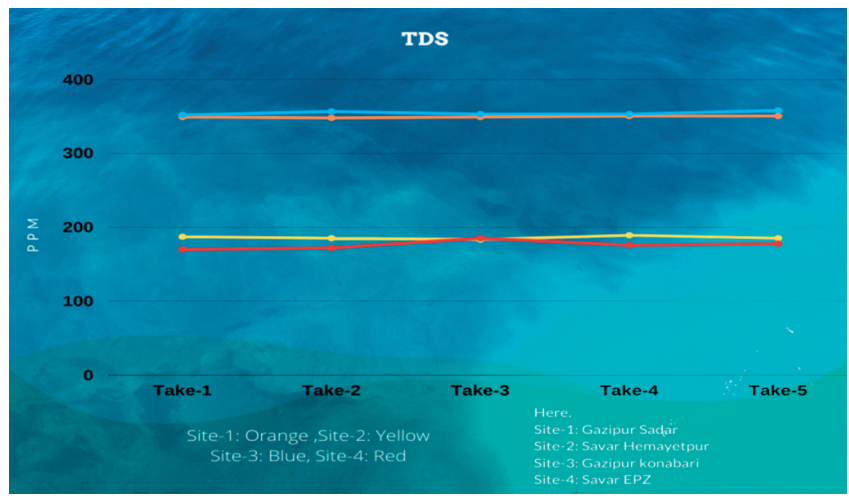

Figure 18: Line chart of TDS values for different takes.

producing knit denim causes a high turbidity value in the wastewater. The wastewater samples in Sites 2 and 3 have less turbidity as the industries located in those areas use fewer colorful chemicals for dyeing.

*3.3.4. TDS Analysis.* TDS values represent the total amount of solids dissolved in the solutions. Table 6 shows the TDS values for different samples collected from different sites. In this case, too, the waste water samples were collected from the same industrial areas mentioned in the previous cases. From Table 6, it can be seen that the TDS value varies a lot from one site to the other. In all cases, high TDS values are very hazardous for water life. The value of TDS varies from an alarming range of 170 ppm–360 ppm. In the experiment, the lowest value of TDS was found to be 170.12 ppm, which is still very high and can cause a lot of damage to the quality of the water. The highest value (358.17) of TDS was noticed in site 3. Figure 18 presents the TDS values for all sites using a line graph for a clearer view. In this study, it was noticed that the factors determining the water quality for industrial areas are higher in most cases compared to the values for normal water. It is vital to control this alarming situation. Garments factory management should manage their waste materials and wastewater released from the garments in a proper way. This data and study will be very helpful for government authorities to monitor and take necessary action for proper waste management. Otherwise, it is becoming more hazardous for people, aquatic species, and the environment. We need to save our environment for the benefit of all of us.

The industries located at sites 1 and site 3 extract more solid-filled wastewater than any other site which resulted in the high value of TDS in the line chart. That happens because of the heavy use of various chemicals in the production system. Those chemicals dissolve in the wastewater and cause a shift in the result bar.

## 4. Conclusion and Future Work

The results and analysis of the IoT-based water quality assessment system for industrial wastewater were presented in this research work. The proposed water quality assessment system can monitor pH, temperature, turbidity, and TDS of water in real-time. Using this system, people can easily monitor water quality in real-time and take necessary steps to control the water quality. The system is also cost-effective and user-friendly. The developed system has TDS, turbidity, pH, and temperature measuring units that are interfaced with several sensors successfully. Our device measures the pH, temperature, turbidity, and TDS of water every 5 sec and shows real-time data on a smartphone screen. After developing the system, it has been tested in real-life context. The water quality factors of the wastewater of four different industrial locations in Bangladesh have been tested. Results and analyses show that industrial waste pollutes the water to a great extent, which needs to be monitored. An IoT-based water quality measurement system can play a very significant role in a Bangladeshi context. Here we are referring to Bangladesh specifically because we know the poor condition of Bangladesh. There are many garments factories in Bangladesh. Most of them stand on the riversides in the country. That is why a large number of rivers are polluted. Proper action should be taken by the proper authorities to purify the water. But it is high time we need to take practical steps. First of all, people need to be aware. We believe that this research will help many industries keep the environment safe. We also hope that people will use this proposed system for the sake of the nation. Using this system, the water quality can be evaluated before supplying it to the masses. The proposed system, research results, and analysis will help to save our water from pollution, and thus improve the health of all living creatures. IoT is a very important emerging technology in healthcare, and will help to increase the life expectancy of people and other living creatures.

The application of this proposed system can be extended further. In future, more sensors can be added to measure other parameters of the water. The system can monitor other factors that pollute the environment by adding more specific sensors to the system.

## Data Availability

No data were used to support the findings of this study.

## Conflicts of Interest

The authors declare that they have no conflicts of interest to report regarding the present study.

## Acknowledgments

The authors would like to thank Taif University Researchers Supporting Project number (TURSP-239), Taif University, Taif, Saudi Arabia.